\documentclass[runningheads]{llncs}
\usepackage{cite}
\usepackage{amsmath,amssymb,amsfonts}
\usepackage{textcomp}
\usepackage{xcolor}
\usepackage{graphicx}
\usepackage{caption}
\usepackage{subcaption}
\usepackage[linesnumbered,ruled,vlined]{algorithm2e}

\newcommand\new[1]{\textcolor{black}{#1}}

\begin{document}
\title{Decentralised Blockchain Management Through Digital Twins}
\titlerunning{Decentralised Blockchain Management Through Digital Twins}
%
\author{Georgios Diamantopoulos \inst{1,2} \and
Nikos Tziritas \inst{3} \and
Rami Bahsoon \inst{2} \and
Georgios Theodoropoulos \thanks{Corresponding author} \inst{4,1}
}
\authorrunning{G. Diamantopoulos et al.}
%
\institute{
Southern University of Science and Technology, Shenzhen, China \and
University of Birmingham, Birmingham, UK \and
University of Thessaly, Lamia, Greece \and
Research Institute of Trustworthy Autonomous Systems, Shenzhen, China 
}
\maketitle              
\begin{abstract}
The necessity of blockchain systems to remain decentralised limits current solutions to blockchain governance and dynamic management, forcing a trade-off between control and decentralisation. In light of the above, this work proposes a dynamic and decentralised blockchain management mechanism based on digital twins. To ensure decentralisation, the proposed mechanism utilises multiple digital twins that the system's stakeholders control. To facilitate decentralised decision-making, the twins are organised in a secondary blockchain system that orchestrates agreement on, and propagation of decisions to the managed blockchain. This enables the management of blockchain systems without centralised control. A preliminary evaluation of the performance and impact of the overheads introduced by the proposed mechanism is conducted through simulation. The results demonstrate the proposed mechanism's ability to reach consensus on decisions quickly and reconfigure the primary blockchain with minimal overhead.

\end{abstract}

\section{Introduction}

As blockchain matures, increasingly sophisticated applications are proposed to leverage its decentralisation, transparency, and security properties. With the advent of permissioned blockchain, the technology has found novel use cases beyond finance, providing the infrastructure for storing traceable and immutable records serving applications in IoT, supply chain management, e-governance, and more \cite{amiri2021permissioned}. However, as the technology matures and blockchain applications evolve and scale, managing blockchain systems becomes challenging.

Blockchains are complex and dynamic systems that experience varying over their lifetime. Consequently, blockchain systems that are unable to adapt to such conditions dynamically are likely to experience periods of inefficiency. This strongly motivates dynamic blockchain management and optimisation. However, blockchain systems rely on their decentralisation property to provide their core features (immutability, transparency and security) necessary for trust. This introduces a novel constraint to managing blockchain systems, maintaining decentralisation, which is not satisfied by traditional approaches \cite{lumineau2021blockchain}. 

\begin{figure}[t]
    \centering
    \includegraphics[width=1\textwidth]{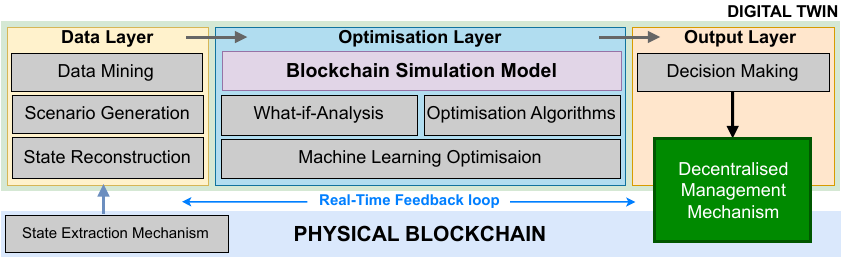}
    \caption{Proposed digital twin architecture \new{as conceptualised} in \cite{diamantopoulos2022digital}}
    \label{fig:digital-twin-arch}
    \vspace{-.4cm}
\end{figure}

Digital Twins\cite{arup} are identified as a promising technology for the problem of dynamic blockchain management and governance. In previous work, a dynamic digital twin architecture for managing blockchains was proposed, and a preliminary evaluation of its ability to manage blockchains was conducted \cite{diamantopoulos2022digital, diamantopoulos2022dynamic}. 
 A high-level architecture for blockchain digital twins is presented in Figure \ref{fig:digital-twin-arch}.  The system relies on a feedback loop between the blockchain and its Digital Twin. In  \cite{diamantopoulos2024dynamic} the physical-to-virtual  direction of the loop was considered, namely extracting and synchronising the state of blockchain networks to update the model within the Digital Twin.  
 
 This paper focuses on closing the feedback loop, proposing a mechanism to implement the virtual-to-physical direction, updating the blockchain system based on the output of the Digital Twin. A core challenge of the above is ensuring that the mechanism does not introduce any centralisation into the blockchain system, as that could result in catastrophic damage. To this end, a mechanism utilising multiple digital twins controlled by the system's stakeholders is proposed. To facilitate agreement, the twins are organised in a secondary blockchain network, which orchestrates decentralised agreement on management decisions and their propagation to the managed blockchain system. The former is enabled by the consensus process and the latter through the resulting blockchain structure. Leveraging the power of digital twins organised in a blockchain network allows for granular control while maintaining decentralisation.
 
The paper structure follows: Section \ref{sec:Solution} presents the challenges and proposed solution to decentralised blockchain management. Section \ref{sec:Exp} presents the experimental setup and the evaluation results. Section \ref{sec:RW} presents the relevant literature while Section \ref{sec:conc} concludes this paper.

\section{A `Blockchain-based` Approach for Decentralised Blockchain Management}
\label{sec:Solution}

\begin{figure}
    \centering
    \includegraphics[width=1\linewidth]{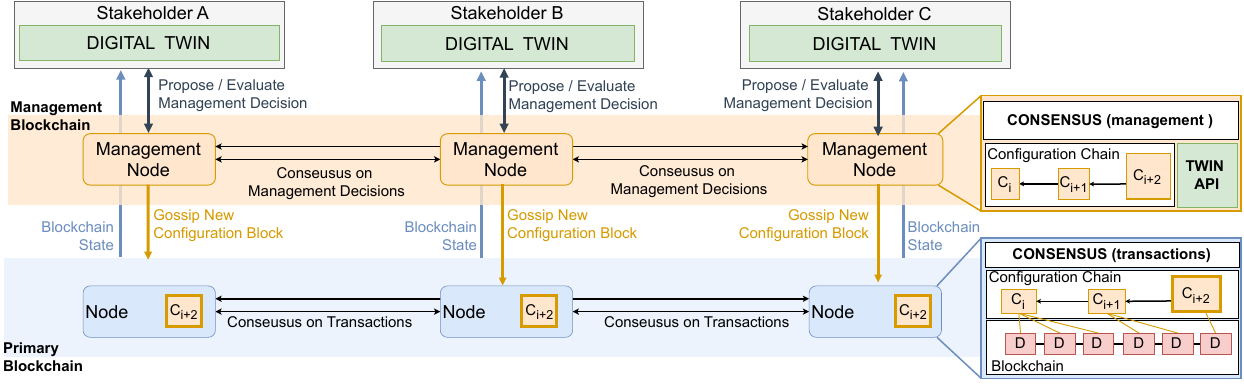}
    \caption{A generic blockchain system managed by the proposed mechanism}
    \label{fig:proposed-architecture}
    \vspace{-.4cm}
\end{figure}

Any system designed to manage blockchains must ensure that the decentralisation of the managed system is not violated. In the context of digital twin-based blockchain management, this manifests as the problem of who controls the digital twin. Blockchain systems provide a decentralised, trusted platform to enable cooperation between a set of stakeholders. Thus, the stakeholders of the blockchain system must collectively steer the management process. This necessitates multiple digital twins to represent the system holders. However, the introduction of multiple twins gives rise to two core challenges: \textbf{i)} How can agreement on a global decision be reached while simultaneously tolerating possible Byzantine behaviour (a core assumption of blockchains)? \textbf{ii)} How can the agreed-upon decisions be applied to the primary blockchain in a decentralised manner while ensuring the resulting blockchain structure can remain verifiable?

A notable observation, that forms the basis of this work, is that blockchain systems are designed to address this set of problems. Blockchain enables decentralised Byzantine fault tolerance consensus while providing a traceable, immutable, and verifiable ledger that tracks the system's evolution. Inspired by the above, this work proposes a `blockchain-based' approach to blockchain management that utilises a secondary blockchain system. Through this, the stakeholder-controlled digital twins can agree upon and propagate management decisions to the primary blockchain system. Specifically, the proposed architecture targets consortium blockchain systems \cite{dib2018consortium}; permissioned blockchains instantiated by mutually untrusting entities to facilitate their cooperation.

\subsection{Agreement}

To ensure decentralised management, multiple twins representing the stakeholders of the primary blockchain must participate in the process, which creates an agreement problem. The proposed solution addresses this by organising digital twins in a secondary blockchain network called the 'management blockchain', which comprises 'management nodes'. Within this system, the management nodes reach consensus on management decisions for the primary blockchain through the blockchain's consensus process. Drawing a parallel to traditional blockchain systems, the digital twins act as the application layer, while the management nodes provide them access to the blockchain network, enabling participation in the consensus process. Figure \ref{fig:proposed-architecture} depicts a generic blockchain system managed by the proposed mechanism.

\begin{algorithm}[t]
    \caption{Management Node Logic}
    \label{alg:agreement}
        \KwData{management node $n$, configuration block interval $CBI$, current cycle $r$}
        \Begin{     
            calculate proposer for $r$ \\
            \If{$n$ is proposer}{
                create and propose configuration block $CB_r$ for cycle $r$ \\
                \If{consensus reached}{
                    gossip signed $CB_r$
                }
                \Else{
                    begin cycle $r+1$
                }
            } \Else{
                wait for proposal ($CB_r$) or cycle timeout \\
                \If{$CB_r$ is a valid under $r$}{
                    validate configuration included in $CB_r$ and cast appropriate vote \\
                }\Else{
                    begin cycle $r+1$
                }
            }
        }   
\end{algorithm}

To arrive at global decisions, a proposer digital twin computes and proposes a decision to the management blockchain through its local management node. This initiates a consensus round with the remaining twins acting as validators. Specifically, the proposer management node packs the decision of its digital twin in a configuration block before proposing it to the system. When acting as validators, management nodes use their respective digital twin to validate the proposed configuration blocks. Since blockchain consensus protocols are agnostic to the content of the blocks, any off-the-shelf protocol can be used. \new{The use of Byzantine Fault Tolerant (BFT) protocols ensures that Byzantine behaviour from the twins can be tolerated.}

The above process is governed by the configuration block interval ($CBI$) hyperparameter, which defines the minimum interval between two subsequent configuration blocks. This provides a mechanism to control rapid fluctuations between configurations, specifically designed to limit fluctuations under unstable conditions in the primary blockchain.

To fairly select proposers, a deterministic Tendermint-style proposer selection algorithm is used \cite{buchman2016tendermint}. Given the hash of the latest configuration block (denoted as $h_l$), the next proposer can be computed as $p_{next} = h_l \mod |MN|$ where $|MN|$ is the number of management nodes. To handle offline or non-quorum proposers, the process is organised into proposal cycles (propose + vote). The agreement process triggered by the $CBI$ allows up to $|M_N|$ cycles. The proposer selection is modified to $p_{next}=(h_l+FC) \pmod{|M_N|}$, where $FC$ is the number of failed cycles since the $CBI$. This guarantees that if a quorum exists, a live quorum member will get to propose a configuration block. Algorithm \ref{alg:agreement} presents the agreement process in a structured manner.

\vspace{.1cm}

\noindent \textbf{Configuration Block Structure:} The configuration blocks of the management blockchain contain system configurations. Specifically, a configuration block at height $i$ ($CB_i$) contains the following: 
\[ 
CB_i = \{H_i, H_{i-1}, P, C^{prim}, proof, height\}
\]
where $H_i$, $H_{i-1}$ are the current and previous configuration block hashes, $P$ is the block proposer, $C$ is the proposed configuration, and $proof$ a proof that consensus was reached. The definition of a configuration $C$ is specific to the application and therefore outside the scope. \new{In general, the configuration of a blockchain system refers to configurable parameters or settings that define its operation. Examples include consensus mechanisms and their parameters (e.g., block interval, validator set), peer discovery, node roles, smart contract upgrade policies, and network topology, among other aspects.}

\begin{algorithm}[t]
    \caption{Node reconfiguration}
    \label{alg:reconfiguration}
        \KwData{node index $i$, local management chain $MC_i$, height of active configuration block $H^i_{cc}$, gossiped configuration block $CB_{new}$, proposed data block $B$}
        \Begin{      
            \While{active consensus round}{
                \If{received valid $CB_{new}$}{
                    gossip $CB_{new}$ to peers\\
                    \If {$CB_{new}$ is future} {
                        request missing blocks \\
                    }\Else{
                        append $CB_{new}$ to $MC_i$
                    }
                }
                \If {configuration of $B \neq H^i_{cc}$}{
                    $B$ is invalid
                }
            }
            after consensus round ends (success or fail) \\
            wait for any missing blocks \\
            \If{new configuration blocks in $MC_i$}{
                update to configuration in latest block\\
                $H^i_{cc} \gets MC_i[last]$.height
            } 
            start next round according to $H^i_{cc}$
        }
\end{algorithm}

\subsection{Reconfiguration}

Once agreement is reached, the scope shifts to the primary blockchain, where the agreed-upon configuration must be verified by the primary nodes and applied. To allow this, the primary blockchain nodes participate in the management blockchain as observer nodes; nodes that do not participate in the consensus process but are aware of the management nodes and keep an up-to-date copy of the blockchain structure (see Figure \ref{fig:proposed-architecture}). Giving primary nodes the ability to see and verify configuration blocks enables the configuration of the primary blockchain to be controlled by the management blockchain structure. Specifically, the latest configuration block controls the active configuration of the primary blockchain. To avoid new configuration blocks from disrupting ongoing consensus in the primary blockchain, reconfiguration is only allowed between consensus rounds. Algorithm \ref{alg:reconfiguration} showcases the reconfiguration logic.

\subsection{Verifiability}

\begin{figure}[t]
    \centering
    \begin{subfigure}[b]{0.43\textwidth}
        \centering
        \includegraphics[width=\textwidth]{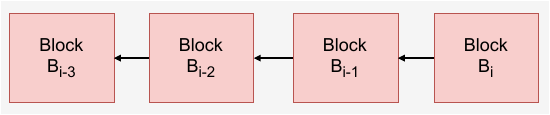}
        \caption{Traditional Blockchain Structure}
        \label{sub:trad-struct}
    \end{subfigure}
    \begin{subfigure}[b]{0.56\textwidth}
        \centering
        \includegraphics[width=\textwidth]{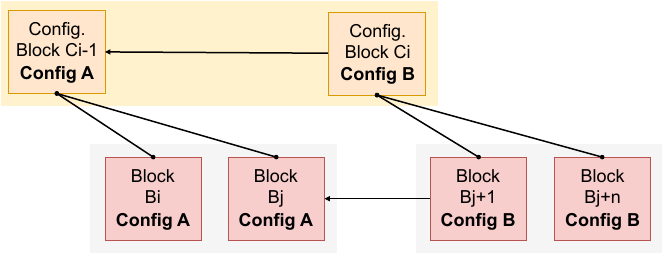}
        \caption{Managed Blockchain structure}
        \label{sub:reconf-struct}
    \end{subfigure}
    
    \caption{Differences between a traditional blockchain structure and the proposed digital twin-managed blockchain structure.}
    \label{fig:blockchain-structures}
    \vspace{-.4cm}
\end{figure}

A fundamental blockchain property and an enabler of blockchain's trust model is verifiability. Blockchain systems achieve verifiability by maintaining a decentralised, immutable, and back-linked list of blocks that contain cryptographic signatures and proofs. This enables nodes to individually verify the system's state by rebuilding it block by block, checking that each block includes valid signatures and a correct reference to the latest block. 

In a reconfigurable blockchain system, the configuration details must be known to allow for validation (active validator list, consensus protocol, block size, etc.). The management blockchain offers a natural solution to the problem of block verification, as all past system configurations are stored in an equally immutable and verifiable blockchain structure. Linking transaction blocks to configuration blocks ensures the configuration details remain immutably linked to the block, allowing future verification. Figure \ref{fig:blockchain-structures} showcases the differences between a traditional blockchain structure and the proposed structure. Furthermore, the links between transaction blocks and the management blockchain structure enable nodes to validate whether newly proposed transaction blocks adhere to the latest configuration. 

\section{Experimental Evaluation}
\label{sec:Exp}

The conducted evaluation aims to measure the sensitivity of management delays and reconfiguration overheads to varying conditions in the primary and management blockchain systems. These factors determine how quickly decisions are taken and the impact of applying such choices to the primary blockchain. Simulation was used to provide the experimental platform for the above evaluation. The proposed mechanism was implemented by modifying the SymbChainSim blockchain simulation \cite{diamantopoulos2025symbchainsim, diamantopoulos2023symbchainsim}. Specifically, an API was implemented to allow the simulated management blockchain to interface with the simulated primary blockchain. The API allowed the primary blockchain to request the simulation of a consensus round in the management blockchain and receive the resulting configuration block along with its associated delay. A gossip event containing the new configuration block is then injected into the primary blockchain simulation at the appropriate time to simulate the propagation of the configuration block along the primary blockchain nodes.

\vspace{.2cm}

\noindent\textbf{Experimental Set-up:}
The experimental setup consists of a primary blockchain system comprising $32$ nodes connected to $16$ random peers over network links, forming a peer-to-peer (P2P) network. The bandwidths of the links are dictated by a normal distribution with $\mu=10MB/s$ and $\sigma=0.5MB/s$ (sampled per message), and the latency is dictated by their geographical distance (calculated based on \cite{goonatilake2012modeling} and randomly assigned geographical locations). The management blockchain configuration consists of $16$ nodes randomly drawn from the $32$ primary blockchain nodes at the start of the simulation. The consensus protocol utilised by both blockchain networks is PBFT. The maximum allowed block size for transaction blocks is $1MB$, and the block interval is set to $0.1s$. The configuration block size is $0.25MB$. The $CBI$ was drawn from a normal distribution ($\mu=30s, \sigma=5s$) over $1$ hour-long simulations. To eliminate noise, the produced configuration blocks always contained the same configuration. In the following experiments, only the parameter under examination varies. \new{This baseline configuration serves as a constant reference for the observation of relative fluctuations in response to changes in system conditions.}

\subsection{Evaluation of the Management Delays}

\begin{figure}
    \centering
    \begin{subfigure}[t]{0.48\textwidth}
        \centering
        \includegraphics[width=\textwidth]{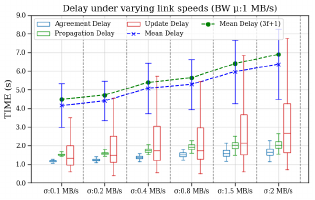}
        \caption{Network link speed ($\mu=1MB/s$)}
        \label{sub:d_net_1}
    \end{subfigure}
    \begin{subfigure}[t]{0.48\textwidth}
        \centering
        \includegraphics[width=\textwidth]{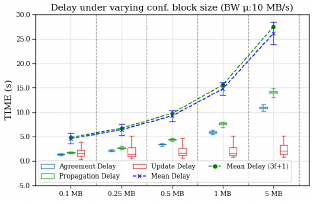}
        \caption{Configuration block size}
        \label{sub:d_conf_b_size}
    \end{subfigure}
    
    \begin{subfigure}[t]{0.48\textwidth}
        \centering
        \includegraphics[width=\textwidth]{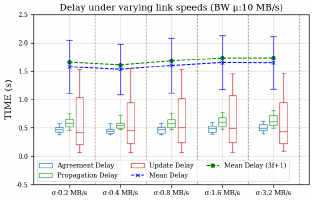}
        \caption{Network link speed ($\mu=10MB/s$)}
        \label{sub:d_net_10}
    \end{subfigure}
    \begin{subfigure}[t]{0.48\textwidth}
        \centering
        \includegraphics[width=\textwidth]{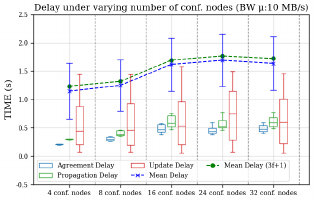}
        \caption{Number of configuration nodes}
        \label{sub:d_num_conf}
    \end{subfigure}

    \caption{Evaluation of management delays}
    \label{fig:delay}
    \vspace{-.4cm}
\end{figure}

The management delay measures the interval between the $CBI$ and the resulting management decisions being reflected in the primary blockchain. In the proposed mechanism, this delay can be broken down into three phases: agreement, propagation, and update. Agreement delay refers to the time required for the management nodes to reach a consensus on a decision. Propagation delay refers to the time needed for the produced configuration block to arrive at the primary blockchain system nodes. Finally, update delay refers to the time between a primary node receiving a new configuration block and the time required to update its local configuration to reflect the contents of that configuration block. The propagation and update phases need only take place on $2f+1$ nodes for the system to start functioning under the new configuration. Consequently, the following results present the delays as observed on the fastest $2f+1$ nodes for each configuration block.

Figures \ref{sub:d_net_1} and \ref{sub:d_net_10} show that the impact of varying link bandwidth is pronounced when the mean link bandwidth is low, with the effects concentrated on update delays. This phenomenon is attributed to the fact that high variation in link performance leads to uneven dissemination of information, that is, some nodes receive information faster than others. This effect can compound and result in a `split' network where nodes with faster links update their configuration earlier than others. The effect was not observed in the case of higher mean bandwidth (Figure \ref{sub:d_net_10}), which diminishes the impact of variations. 

Finally, figures \ref{sub:d_conf_b_size} and \ref{sub:d_num_conf} evaluate the effect of parameters related to the management blockchain. The results show that the size of the configuration block has a pronounced effect on the agreement and propagation delay, but has little impact on the update delay. This is expected as larger blocks increase the size of the network messages that are necessary to facilitate agreement and, of course, propagate the block to the primary blockchain. The number of configuration nodes, on the other hand, has little effect on the reconfiguration delay. This is a positive observation as systems with numerous stakeholders can adopt this approach without worry about increased decision overheads.

\subsection{Evaluation of the reconfiguration overheads}

\begin{figure}
    \centering
    \begin{subfigure}[t]{0.48\textwidth}
        \centering
        \includegraphics[width=\textwidth]{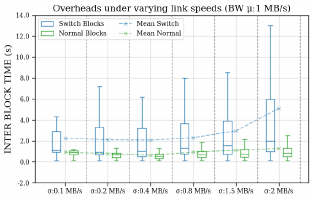}
        \caption{Network link speed ($\mu=1MB/s$)}
        \label{sub:o_net_1}
    \end{subfigure}
    \begin{subfigure}[t]{0.48\textwidth}
        \centering
        \includegraphics[width=\textwidth]{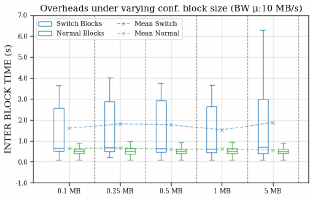}
        \caption{Configuration block size}
        \label{sub:o_conf_b_size}
    \end{subfigure}
    
    \begin{subfigure}[t]{0.48\textwidth}
        \centering
        \includegraphics[width=\textwidth]{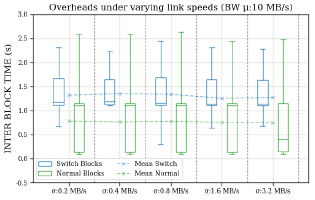}
        \caption{Network link speed ($\mu=10MB/s$)}
        \label{sub:o_net_10}
    \end{subfigure}
    \begin{subfigure}[t]{0.48\textwidth}
        \centering
        \includegraphics[width=\textwidth]{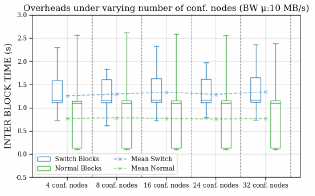}
        \caption{Number of configuration nodes}
        \label{sub:o_num_conf}
    \end{subfigure}

    \caption{Evaluation of the impact on the primary blockchain (overhead)}
    \label{fig:overheads}
    \vspace{-.4cm}
\end{figure}

Besides the management delay, another critical factor to consider is the impact of the dynamic management process on the primary blockchain system. Due to the nature of reconfiguration, overheads are contained in the interval between two sequential transaction blocks produced under different configurations. Intuitively, this is the case because blocks produced before that were created under the old configuration, and later blocks will be made under the new configuration. Thus, measuring the overhead of reconfiguration can be effectively captured by comparing the inter-block times between sequential blocks under the same configuration (normal blocks) to those of sequential blocks under different configurations (switch blocks). The observed differences in duration constitute the reconfiguration overhead.

The reconfiguration overheads appear to be less sensitive to the same system parameters as compared to the management delays (Figures \ref{sub:o_net_1}, \ref{sub:o_net_10}, \ref{sub:o_conf_b_size}, \ref{sub:o_num_conf}). This observation suggests that the effects of reconfigurations have minimal impact on the operation of the primary blockchain system. It is worth noting that the above overheads are in the context of the $CBI$, meaning they are only incurred once each time the management blockchain produces a new configuration block. Another interesting phenomenon is that, in most observations, the median switch block delay is close to the median delay of normal blocks (denoted by the bar inside the box component of the box plots). This demonstrates that most reconfiguration operations added minimal overheads, with the observed difference in the mean being attributed to outliers. This behaviour can be exploited to allow timing reconfiguration operations to minimise overhead.

\section{Related Work}
\label{sec:RW}

The problems of dynamic blockchain management and governance have been gaining traction in the blockchain literature. As dynamic systems, blockchains stand to benefit significantly from dynamic optimisation as it directly leads to improved performance and quality of service. Furthermore, as blockchain establishes itself as a mainstream technology, combined with its integration in critical applications, the need for novel governance mechanisms suitable for blockchain systems has become increasingly vital.

\vspace{0.1cm}

\noindent\textbf{Dynamic Blockchain Optimisation:} One of the most influential works in dynamic blockchain optimisation/management is \cite{rl}, which proposes a reinforcement learning-based approach for dynamic adaptation of a blockchain system's consensus protocol, block size and block interval to improve performance under varying system conditions. However, despite the impressive results exhibited by the above mechanisms, the challenge of integrating these dynamic decisions into the blockchain system in a decentralised manner was not addressed. Furthermore, previous work \cite{diamantopoulos2022digital, diamantopoulos2024dynamic}, conducted by the authors, has also showcased the performance benefits of a digital twin-managed blockchain, including improved transaction latency, transaction throughput, and block times. However, these experiments were also performed under a centralised management model. In conclusion, the challenge of decentralised blockchain management remains an open problem in the blockchain literature and constitutes the main challenge addressed by this work. 

\vspace{.1cm}

\noindent\textbf{Decentralised Blockchain Governance:}
Updating blockchain systems in a way that maintains decentralisation is also a core challenge in the blockchain governance literature. Current approaches to blockchain governance can be categorised into two major categories: on-chain and off-chain (these approaches are explained in detail in \cite{reijers2021now}). On-chain approaches rely on hard-coded rules within the blockchain network, typically in the form of smart contracts. Although decentralised, on-chain solutions offer limited control. On the other hand, off-chain solutions rely on external platforms, through which the community system stakeholders can propose and discuss changes. Although off-chain approaches can control most aspects of a blockchain system, the decentralisation of off-chain solutions is often poor and hard to verify. Furthermore, agreed-upon changes are slow to reflect in the blockchain system. To this end, the proposed mechanism can offer both the flexibility of off-chain approaches, with the decentralisation guarantees of on-chain solutions, while additionally enabling automated decision-making powered by the digital twin.

\vspace{.1cm}

\noindent\textbf{Multiple blockchain structures:} 
Ideas utilising multiple blockchain data structures have been explored in the blockchain literature; however, existing approaches focus on either increasing security by storing sensitive data into a separate blockchain \cite{kan2018multiple} or enhancing the system's performance by balancing transaction processing across multiple chains \cite{sohan2021increasing}. Furthermore, the proposed mechanism fundamentally differs from layer-two protocols \cite{gudgeon2020sok} and side-chains \cite{garoffolo2018sidechains}. Specifically, layer-two protocols explore ways to utilise external mechanisms to improve blockchain systems while relying on the consensus protocol of the primary blockchain system itself. In contrast, the proposed mechanism is a fully featured blockchain network with a separate consensus protocol. On the other hand, side-chains focus on methods for securely sharing value (such as tokens or coins) between existing blockchain networks that serve different underlying applications. To the best of our knowledge, the proposed solution is the first to suggest a secondary blockchain layer utilised for dynamic blockchain management.

\section{Conclusion}
\label{sec:conc}

This work presents a 'blockchain-based' decentralised blockchain management mechanism based on digital twins, completing the feedback loop of blockchain digital twins. The basis of this work is that a blockchain-based approach to blockchain management can inherently satisfy all requirements for decentralised blockchain management. The proposed mechanism thus revolves around a secondary management blockchain network whose participants are the digital twins controlled by the system stakeholders. The management blockchain implements a consensus process to reach agreement on new system configurations driven by the decisions of the digital twins. To reflect these decisions in the primary blockchain, the nodes of the primary blockchain participate as observer nodes in the management blockchain, keeping an up-to-date version of its blockchain structure. Thus, the active configuration in the primary blockchain can be dictated by the latest block in the management blockchain. An experimental evaluation of a simulated blockchain system assessed the sensitivity of reconfiguration delay and overheads to various system conditions. The results of the experimental evaluation showed that the management blockchain can produce configuration blocks quickly, while the overhead of switching configuration observed in the primary blockchain remains minimal.

\vspace{-.1cm}
\section*{Acknowledgments}
\vspace{-.1cm}

This research was supported by: Research Institute of Trustworthy Autonomous Systems, Southern University of Science and Technology, Shenzhen 518055, China; Program for Guangdong Introducing Innovative and Entrepreneurial Teams under Grant 2017ZT07X386; SUSTech-University of Birmingham Collaborative PhD Programme. 

\vspace{-.1cm}
\bibliographystyle{splncs04}
\bibliography{main.bib}

\end{document}